\documentclass[aps, pra,twocolumn,preprintnumbers,floatfix]{revtex4-1}
\usepackage{graphicx}
\usepackage{dcolumn}
\usepackage{bm}
\usepackage{latexsym,epsfig}
\usepackage{graphicx}
\usepackage{verbatim}
\usepackage{comment}
\usepackage{amsmath}
\usepackage{amssymb}
\usepackage{stmaryrd}
\usepackage{color}
\usepackage{epstopdf}
\usepackage{grffile}
\usepackage{float}
\DeclareGraphicsExtensions{.eps}

\usepackage{soul}
\usepackage{ulem}
\newcommand{\beq}{\begin{equation}}
\newcommand{\eeq}{\end{equation}}
\newcommand{\bea}{\begin{eqnarray}}
\newcommand{\eea}{\end{eqnarray}}
\newcommand{\ben}{\begin{eqnarray*}}
\newcommand{\een}{\end{eqnarray*}}
\newcommand{\bfig}{\begin{figure}}
\newcommand{\efig}{\end{figure}}

\usepackage{hyperref}
\hypersetup{
    colorlinks=true,      
    urlcolor=blue,
    citecolor=blue,
    linkcolor=blue
}

\begin{document}
\title{Flux enhanced localization and reentrant delocalization in the quench dynamics of two interacting bosons on a Bose-Hubbard ladder}
\author{Mrinal Kanti Giri$^{1}$}
\thanks{These authors contributed equally to this work}
\author{Biswajit Paul$^{2,3}$}
\thanks{These authors contributed equally to this work}
\author{Tapan Mishra$^{2,3}$}
\email{mishratapan@niser.ac.in}
\affiliation{
$^1$Department of Physics, National Tsing Hua University, Hsinchu 30013, Taiwan\\
$^2$School of Physical Sciences, National Institute of Science Education and Research, Jatni 752050, India\\
$^3$Homi Bhabha National Institute, Training School Complex, Anushaktinagar, Mumbai 400094, India}

\date{\today}

\begin{abstract}
We study the quench dynamics of two bosons possessing onsite repulsive interaction on a two-leg ladder and show that the presence of uniform flux piercing through the plaquettes of the ladder favors the localization of the bound states in the dynamics. We find that when the two bosons are symmetrically initialized on the edge rung of the ladder, they tend to edge-localize in their quantum walk - a phenomenon which is not possible in the absence of flux. On the other hand, when the bosons are initialized on the bulk rung they never localize and exhibit linear spreading in their quantum walk. Interestingly, however, we find that in the later case a finite flux favours localization of the bulk bound states in the presence of sufficiently weak quasiperiodic disorder which is otherwise insufficient to localize the particles in the absence of flux. In both the cases, we obtain that the localization in the dynamics strongly depends on the combined effect of the flux and interaction strengths, as a result which we obtain a signature of re-entrant delocalization as a function of flux (interaction) for fixed interaction (flux) strengths.

\end{abstract}

\maketitle

\section{Introduction}
The dynamics of a quantum states following a sudden quench of a system parameter reveals important insights about the quantum phases, spreading of correlation and entanglement, transport properties, topology and chaos~\cite{ Polkovnikov2011, aditi_rev_2018,Kollath2008,Zoller2012,Zoller2018,Nagerl2013,Barbiero2018,Sadler2006,quench2,niels2007, Aidelsburger2018,Paredes2021,Arnab2021}. Due to the recent experimental progress in observing quench dynamics in artificial systems, a great deal of studies have been performed to understand the non-equilibrium dynamics both theoretically and experimentally~\cite{Bloch2008,Schneider_2012,Ronzheimer2013,Xia2015, Moeckel_2010,Sengupta2011,Pavel2018,Guardado2021,quench3,quench4,quench6,quench7,quench8,Sopena2021}. While the non-interacting systems are easier to handle, many-body effects pose serious constraints in addressing the dynamics of interacting particles in practice. On the other hand the theoretical analysis of such systems requires sophisticated numerical methods. In such a scenario, the quench dynamics of a state with few interacting particles also known as the quantum walk (QW) offers a unique platform to achieve enough insights about the interacting systems. Often such dynamics also reveal novel phenomena which does not occur in the true many-body limit. 

Recently, the QW of interacting particles have been extensively studied to understand the effect of strong correlations, particle statistics, disorder, external gauge field and topology in lattice systems~\cite{lahini2012qw, Greiner_walk,Peruzzo2010, PhysRevA.101.032336,PhysRevE.106.044215, Tai_2017, chalabi2019synthetic,maria_doublon,Chaohong_etc,poulios2014quantum, yoshi2020,mondalwalk,fermi_bose_walk,Xue2015, PhysRevB.86.195414, giri2022nontrivial, Giri2021,Xie2020} . In this context signatures localization transition, topological character,  Bloch oscillation, chiral dynamics, dynamics of quasiparticle excitation have been theoretically studied and experimentally observed~\cite{Xie2019, Xie2020, Hugel2014ChiralLA, maria2022, Gadway2017direct,Kitagawa2012_topological_QW,Padhan2022, Gadway2018a, Meier2018,Greiner_walk, Longhi_BO_2012, cai2021, Johnson2014,sowinski_sarkar,Lebugle2015}. Among these, the simplest yet interesting set up is the dynamics of two interacting particles where the interplay of interaction and nature of energy bands play crucial role in the quench dynamics. One such phenomenon is the signature of repulsively bound pair of strongly interacting particles in one dimensional lattices~\cite{Winkler2006}. For the case of bosons, while for the initial state with two bosons residing on the same site results in a local bound state due to the strong onsite interaction, the two non-local bosons never form bound state due to fermionization. However, strong off-site interaction favors the non-local bound pairs in the dynamics~\cite{bloch_magnon_expt,Luis-Arya} of non-local bosons. On the other hand doublon bound states have been predicted in the dynamics of distinguishable particles~\cite{Dias2007,sowinski_sarkar,giri2022nontrivial,Dias2016}. 

On the other hand, a completely different scenario appears in the case of a two-leg ladder where the dynamics of hardcore bosons exhibits features of bound states along the rung of the ladder. A recent experiment based on superconducting circuit has revealed the edge localization of a bosonic rung-pair in the dynamics of two hardcore bosons initialized on the edge rung of a two-leg Bose-Hubbard ladder with uniform rung and leg hopping strengths~\cite{Ye2019}. However, when in the bulk rung, the bosons exhibit linear spreading in the dynamics implying no localization. In a subsequent theoretical study by {\it{Li et. al.}}~\cite{Fan2020}, it has been shown that the non-trivial edge localization is due to the complete flatness of the  bound state band in the two-particle band structure only in the limit of hardcore onsite interaction.

In this paper however, we show that when a pair of bosons are subjected to an artificial gauge field~\cite{Aidelsburger_rev_gauge_field, Goldman_rev_gauge_field_2014}, they tend to edge localize even in the presence finite onsite interaction. We study the QW of two interacting bosons on a two-leg flux ladder~\cite{dhar2012, dhar2013, lehurflux2013,Tokuno_2014,lehurflux2015,Mishra2016,mishraflux2018, piraud2015vortex,greschner2016,greschner2015,Rashi_2017, Giamarchi2023,Sebastian2015,oktel2015, Buser2020, ceven2022} and examine the interplay between the onsite  interaction strength and the flux piercing through the plaquette of the ladder in the quench dynamics. 
We find that in the absence of flux, in contrast to the hardcore bosons, bosons with finite onsite interaction when symmetrically initialized on the edge-rung of the ladder exhibit linear spreading. However, in the softcore case we find that the onset of flux results in the edge localization of the bosons. Interestingly, this flux enhanced edge localization is found to occur for some intermediate values of interaction strengths, as a result of which we obtain a re-entrant behaviour in the dynamics as a function of interaction when the flux is fixed. 

On the other hand, when the bosons with finite onsite interaction are initialized on the bulk rung of the ladder, they always exhibit linear spreading in the QW. However, in this case we find that in the presence of a sufficiently weak quasiperiodic disorder, the flux favors the localization of the bosons for a range of values of interaction strengths. In both the cases we find that the localization involves the bosonic bound states which is more favourable in the limit of strong rung hopping compared to leg hopping strengths. In the following we discuss these bahaviour in detail. 

\begin{figure}[t]
    \centering
    \includegraphics[width=1\columnwidth]{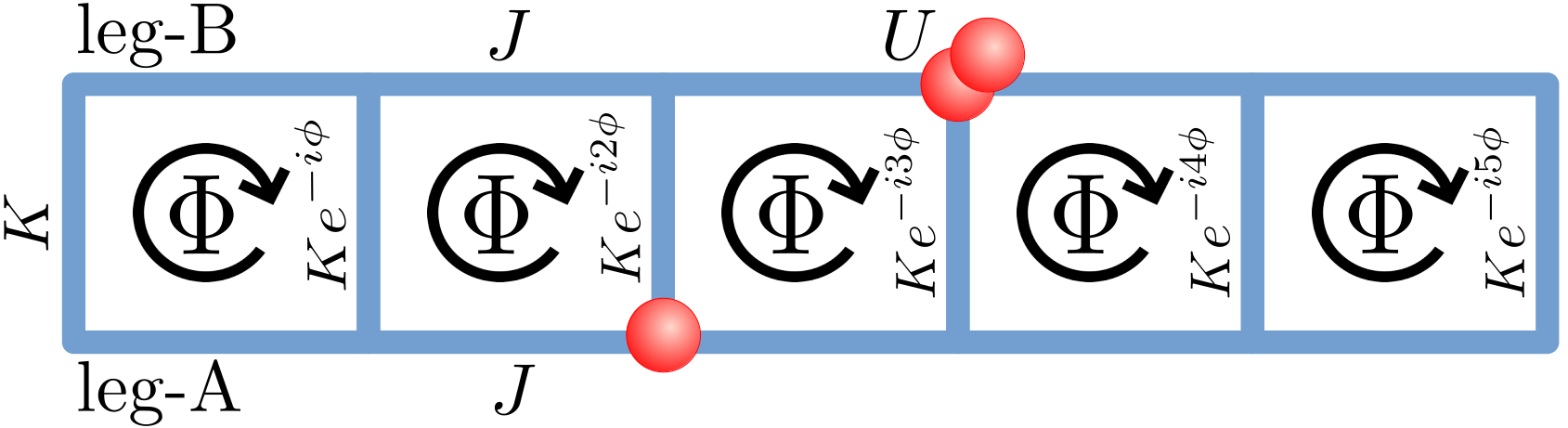}
    \caption{Figure depicts a two-leg Bose-Hubbard ladder in the presence of uniform flux. $J$ and $K$ are the leg and rung hopping strengths respectively. $U$ is the onsite interaction between the bosons and $\Phi$ is the flux piercing through each plaquette and $\phi$ is the phase acquired by the particles. }
    \label{fig:1}
\end{figure}

This paper is structured as follows. In Section II, we present the model of the two-leg ladder subjected to uniform magnetic flux generated by an artificial gauge field for bosons. In Section III, we provide the main results in two parts. In the first part we discuss the edge localization of two interacting bosons and in the second part we discuss the localization in the bulk. 
Lastly, in Section IV, we provide a brief summary of our result.

\section{Model}
We consider the system of interacting bosons on a two-leg ladder in the presence of the uniform magnetic flux as depicted in Fig.~\ref{fig:1}. The Hamiltonian for this setup is given by
\begin{equation}
\begin{split}
    H &= \frac{U}{2}\sum_{l,\sigma} \hat{n}_{l,\sigma}(\hat{n}_{l,\sigma}-1) -J\sum_{l,\sigma} \;({\hat{b}}_{l,\sigma}^{\dagger} {\hat{b}}_{l+1,\sigma}+h.c.)\\ 
    & -\;K\sum_{l}\;( e^{-i l\phi} \;{\hat{b}}_{l,A}^{\dagger} {\hat{b}}_{l,B}  + h.c), 
\end{split} \label{eqn:ham1}
\end{equation}
where $ \sigma \in A,B$ denotes the leg index.  $\hat{b}_{l,\sigma}$ ($\hat{b}_{l,\sigma}^{\dagger}$) are the bosonic annihilation  (creation) operator on rung $l$ of leg-$\sigma$ and $\hat{n}_{l,\sigma}$ is the number operator on rung $l$ of leg $\sigma$. $J$ and $K$ denote the amplitudes of the intra-leg and inter-leg nearest-neighbor hopping respectively. Due to the freedom of the gauge choice in this model, we consider the Peierls phase factors associated with the rungs of the ladder such that 
when a particle encircles a plaquette of the ladder,
its wave function acquires a phase factor $\phi$~\cite{greschner2016, Buser2020}. Here the acquired phase is given as $\phi=\pi\Phi/\Phi_0$, where $\Phi$ is the magnetic flux associated with each plaquette and $\Phi_0=h/e$ is the magnetic flux quantum.

The quantum walk is studied by employing the unitary time evolution protocol as $\left|\Psi(t) \right\rangle = \hat{U}(t) \left|\Psi(0) \right\rangle$ 
where $U(t) = e^{-iH(t)/\hbar}$ and $|\Psi(0)\rangle$ is an initial state. 
For our studies, we consider the initial state by placing two particles on the rung of a ladder which is given by 
\begin{equation}
    |\Psi(0)\rangle= \hat{b}_{l,A}^\dagger \hat{b}_{l,B}^\dagger |vac\rangle.
    \label{eq:initial_state}
\end{equation}
Here, $l$ denotes the rung and $|vac\rangle$ is the vacuum state. The study is performed by exactly solving the Hamiltonian shown in Eq.~\ref{eqn:ham1} for a ladder of length $L = 25$ rungs. For our analysis, we consider stronger rung hopping i.e. $K=5$ and fix $J=1$ which sets our energy scale unless otherwise mentioned.

\section{Results}
In this section we discuss our main findings in detail. We first focus on the edge localization of softcore bosons in the presence of flux. Then we extend our analysis to explore the possibility of the bosons initialized in the bulk of the ladder where we explore the effect of quasi-periodic disorder in the QW. 

\subsection{Edge localization}
\begin{figure}[t]
	\centering
	\includegraphics[width=1\columnwidth]{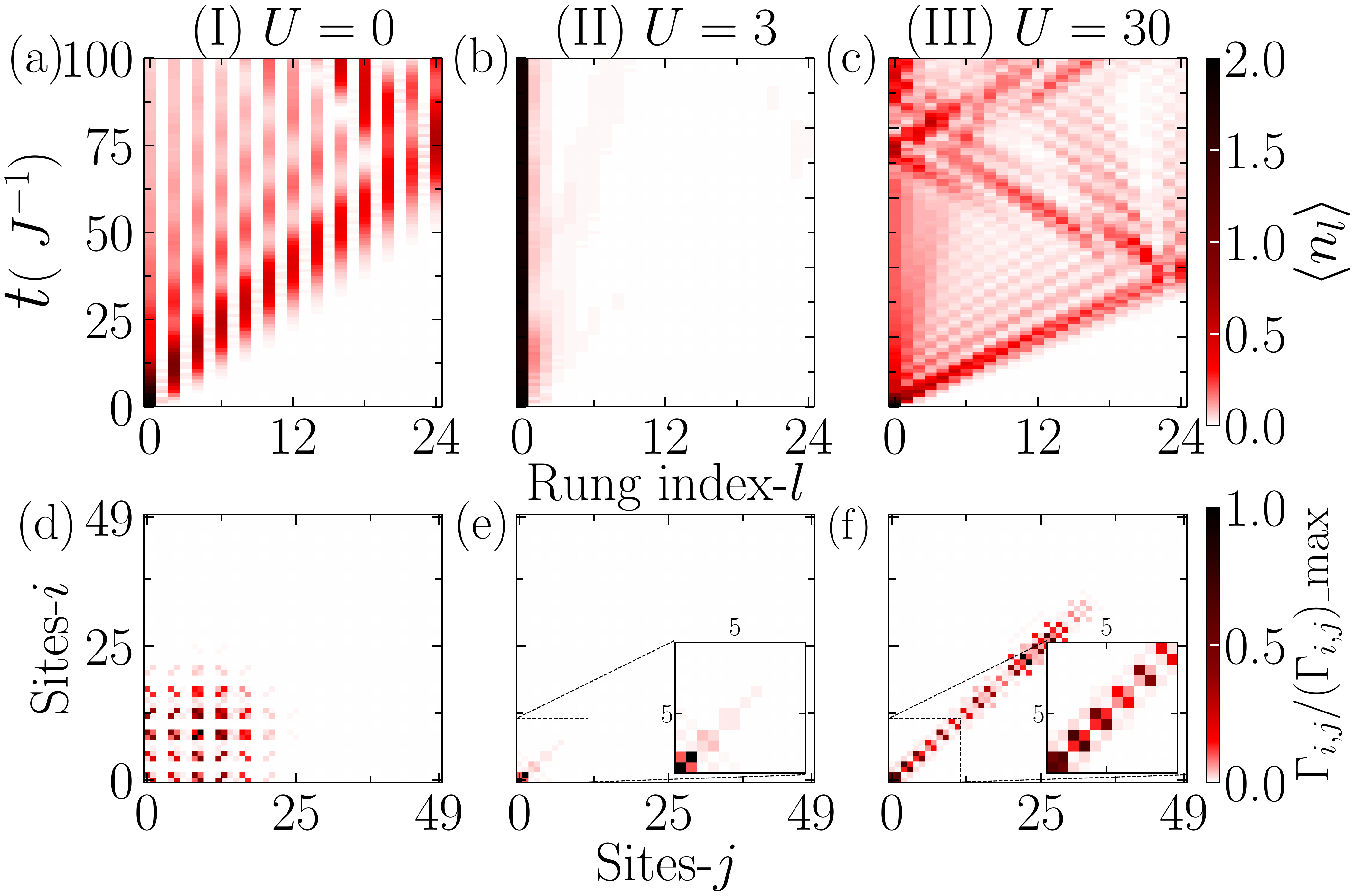}
	\caption{
 (a-c) Show the density evolution of two particles in a two-leg ladder for $U=0,~3$ and $30$ respectively, with the initial states $|\psi(0)\rangle$ given in Eq.~\ref{eq:ini1}. (d-f) Show the  correlation matrix $\Gamma_{i,j}$ at time $t=20(J^{-1})$ for the same values of $U$ as in the upper panel. Here we consider $\phi=\pi$. }
	\label{fig:softcore-density_edge}
\end{figure}

In this subsection we study the QW by considering an initial state $|\Psi(0)\rangle$ given in Eq.~\ref{eq:initial_state} for $l=0$ i.e. 
\begin{equation}
    |\Psi(0)\rangle=\hat{b}_{0,A}^\dagger \hat{b}_{0,B}^\dagger |vac\rangle
    \label{eq:ini1}
\end{equation}
in which the bosons are initialized at the left most edge rung of the ladder. As already mentioned in the introduction, the QW with this initial state for hardcore bosons results in an edge localization of the particles due to the appearance of the flat band corresponding to the rung-pair state. However, for finite onsite interaction $U$ (softcore bosons), the edge localization is forbidden and the bosons perform a linear spreading in the QW. Here, we examine the QW of softcore bosons in the presence of finite flux strength. To this end we first plot the temporal evolution of the onsite particle density given as
\begin{equation}
    \langle \hat{n}_{l}(t)\rangle = \langle \hat{n}_{l,A}(t)\rangle + \langle \hat{n}_{l,B}(t)\rangle ,
\end{equation} 
which is the total density of bosons on a particular rung $l$ of the ladder. 
In Fig.~\ref{fig:softcore-density_edge}(a-c) we show that the spreading of the density for different values of onsite interaction strengths $U$ and $\phi=\pi$. We obtain that when $U=0$, the QW exhibits a faster spreading of the density throughout the lattice as can be seen from Fig.~\ref{fig:softcore-density_edge}(a). However, as $U$ increases and reaches a moderate value i.e. $U\sim3$, a clear localization of the particles at the initial position (i.e. the left edge) is seen in Fig.~\ref{fig:softcore-density_edge}(b).
Unexpectedly, as the interaction strength increases further, the edge localization gradually fades away and a complete delocalization of the particles occurs for very strong interaction, resulting in the linear spreading again as depicted in Fig.~\ref{fig:softcore-density_edge}(c) for $U=30$. 

\begin{figure}[b]
	\centering
	\includegraphics[width=1\columnwidth]{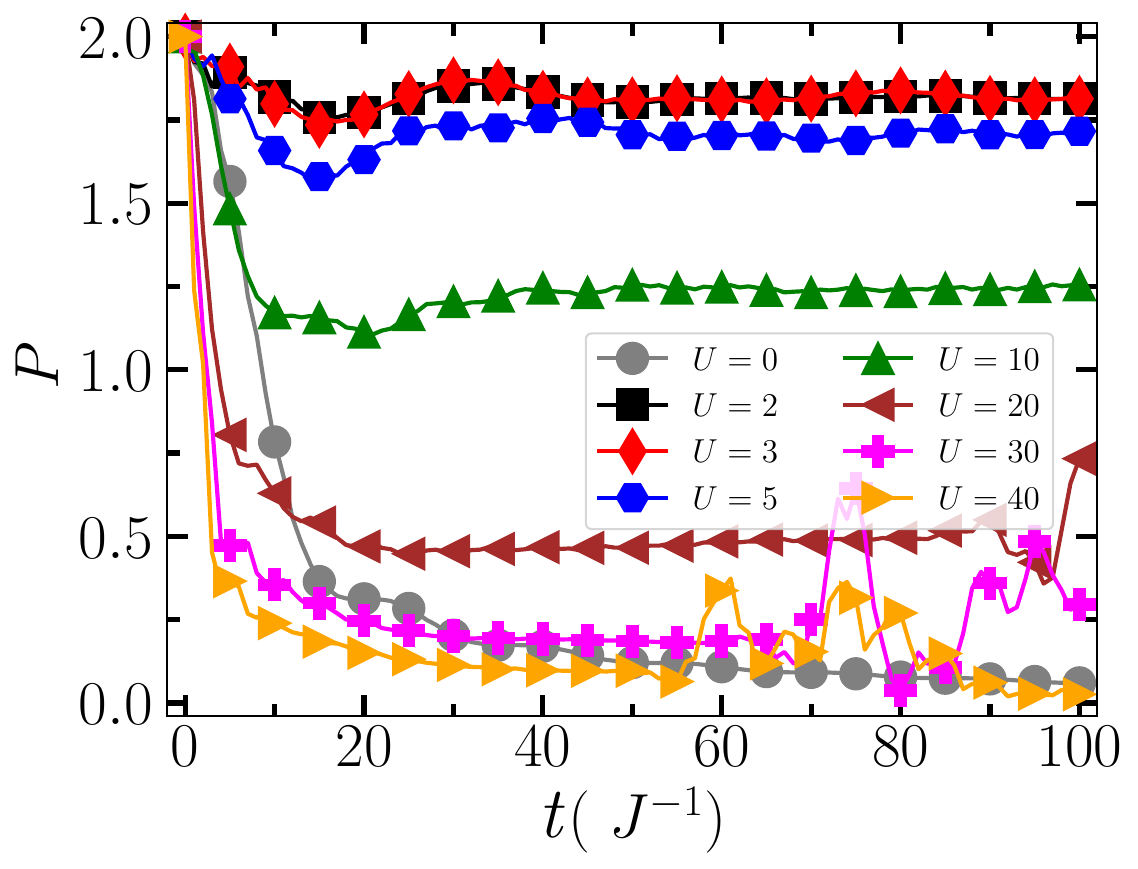}
	\caption{ $P$ as a function of $t(J^{-1})$ for different values of $U$ and for $\phi=\pi$.}
	\label{fig:P(t)_Vs_t_vary_U}
\end{figure}

\begin{figure}[t]
	\centering
    \includegraphics[width=1\columnwidth]{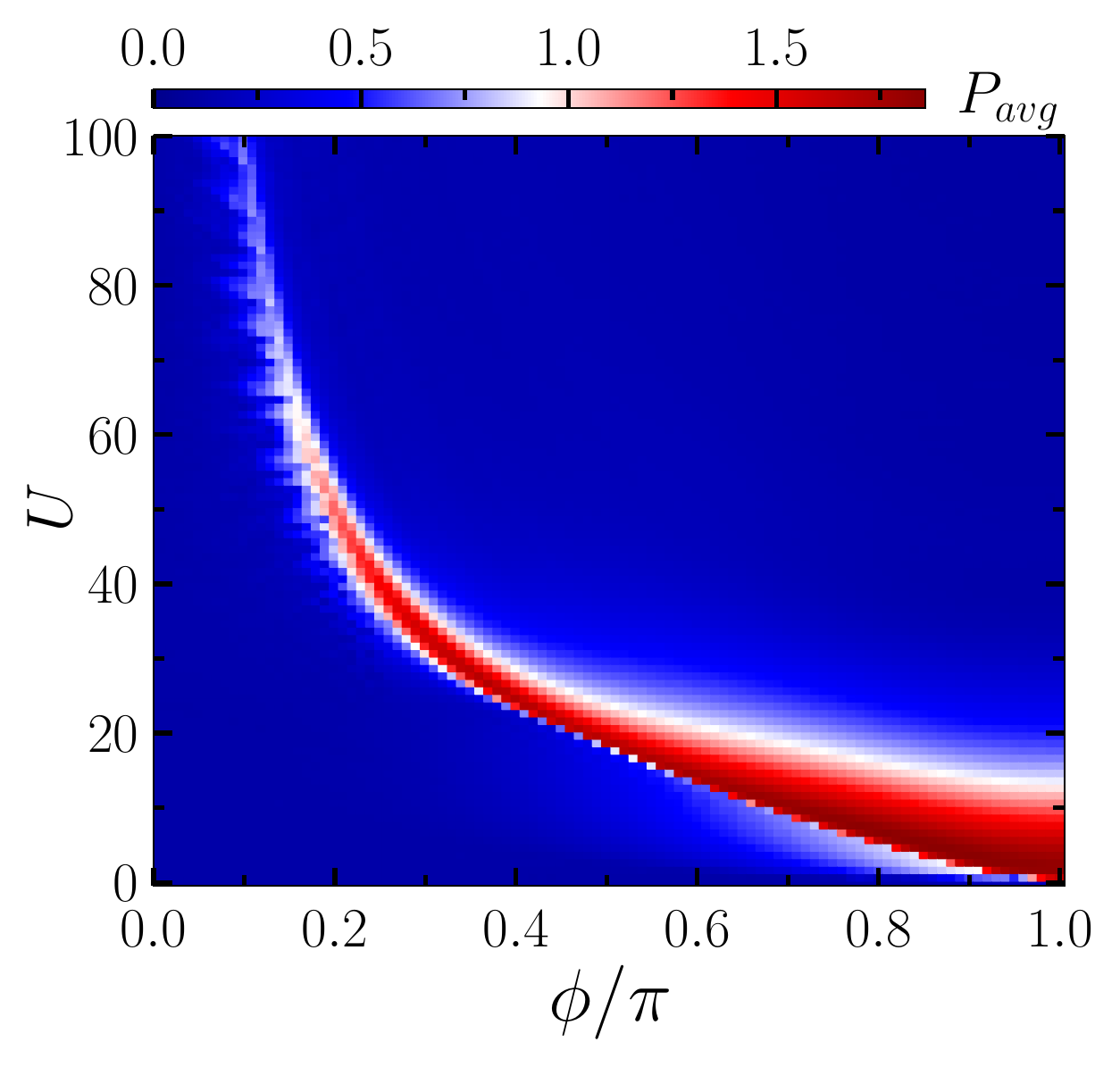}
	\caption{$P_{avg}$ is plotted as a function of $U$ and $\phi/\pi$ for the initial state $|\psi(0)\rangle$ given in Eq.~\ref{eq:ini1}. The red central patch corresponds to the range of $U$ for which the localization is maximum for each values of $\phi$.  Here the $P_{avg}$ is calculated by averaging time between $0$ to $T=1000(J^{-1})$ with a step size of $\Delta t=2(J^{-1})$.}. 
    \label{fig:P_avg_vs_U}
\end{figure}
We quantify this edge localization of bosons by examining the time evolution of the edge-rung density defined as  
\begin{equation}
    P(t) = \langle \hat{n}_{0}(t) \rangle.
\end{equation}
In Fig.~\ref{fig:P(t)_Vs_t_vary_U} we plot $P$ as a function of $t$ for different values of $U$ and $\phi=\pi$. It can be seen that for $U=0$ (grey circles), $P$ saturates to a very small value in time. However, as $U$ increases, $P$ saturates to a finite and large value indicating a maximum probability of occupation on the edge rung which is the signature of an edge localization of the particles. As $U$ increases further, the saturation value of $P$ tends to fall indicating no edge rung localization.  From this analysis we obtain that the saturated value of $P$ is maximum for $U\sim3$ for $\phi=\pi$. 

At this point it is understood that a finite flux favours an edge localization of a rung-pair state of softcore bosons for some particular range of values of $U$. This leads to a re-entrant dynamics that exhibits a delocalization to localization and then to delocalization as a function of $U$. Now the question is how does the range of $U$ for which the edge localization is favored depend on the flux strength? To examine this we calculate the time averaged value of the edge-rung density
\begin{equation}
    P_{avg}=\frac{1}{m}\sum_{t=0}^{T}P(t),
\end{equation}
where $m$ is the number of time steps and plot it as a function of $U$ for $\phi/\pi$ in Fig.~\ref{fig:P_avg_vs_U}. In the figure the blue (red) region corresponds to small (large) values of $P$. It can be seen that as the flux decreases, the range of $U$ at which the edge localization is maximum (red region in Fig.~\ref{fig:P_avg_vs_U}) shift towards the higher and higher values and eventually tends to fade away in the limit of strong interaction and weak  $\phi/\pi$.

In order to discern the states involved in the edge localization we compute the time evolved two-particle correlation function defined as
\begin{equation}
    \Gamma_{i,j}=\langle \hat{b}_i^{\dagger} \hat{b}_j^{\dagger} \hat{b}_j \hat{b}_i \rangle ,
\end{equation}
Where, $\hat{b}_i(\hat{b}_i^\dagger)$ is the particle annihilation (creation) operator and $i,~j$ are the site indices of the ladder. For correlation calculation, the indexing starts
from the left most site of leg-A of the ladder such
that even (odd) indices are on leg-A (leg-B). We plot  $\Gamma_{i,j}$ for three different values of $U$ such as $U=0,~3$ and $30$ at $t=20(J^{-1})$ in the time evolution in Fig.~\ref{fig:softcore-density_edge}(d), (e) and (f), respectively. When $U=0$, the free particle dynamics can be seen as the uniformly distributed correlation matrix elements $\Gamma_{i,j}$ in Fig.~\ref{fig:softcore-density_edge}(d). The asymmetric distribution in the correlation matrix element in this case is due to the location of the initial state. 
\begin{figure}[b]
	\centering
    \includegraphics[width=1\columnwidth]{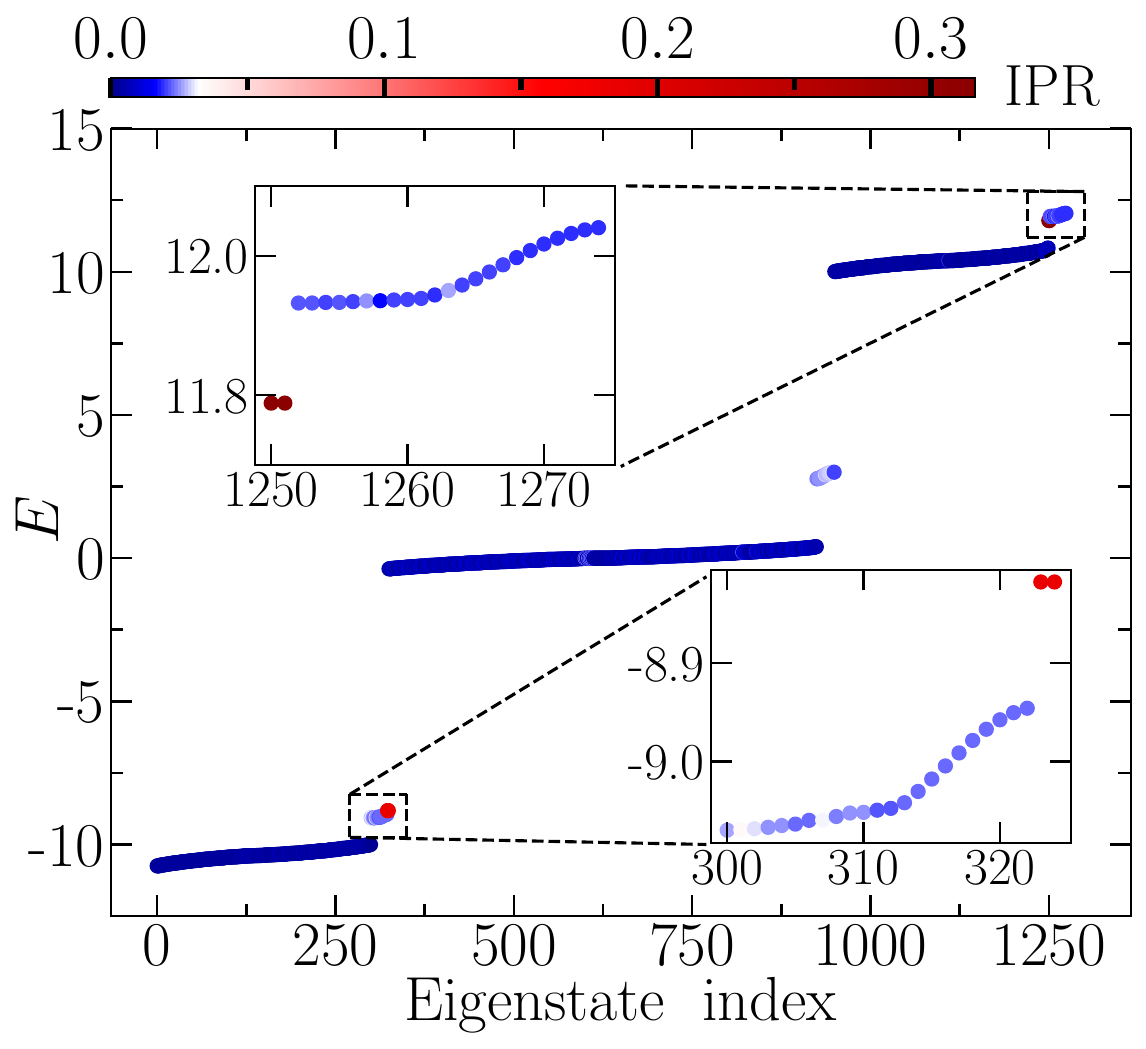}
	\caption{IPR of the states plotted as a function of energy eigenvalues and the eigenstate index for $U=3$ and $\phi=\pi$. }
	\label{fig:Energy_vs_index}
\end{figure}
For $U=3$, only a few finite elements appear at the lower left corner of the correlation matrix $\Gamma_{i,j}$ due to the localization of the particles at the left edge of the ladder (see Fig.~\ref{fig:softcore-density_edge}(e)). A zoomed in picture of these matrix elements (inset of Fig.~\ref{fig:softcore-density_edge}(e)) reveals that the diagonal terms are finite which is an indication of the two-particle onsite bound pair. Apart from the diagonal terms, we also find finite elements one site above or below the diagonal which is the signature of the rung pairs i.e. a bound state of the bosons residing on each sites of a rung. This clarifies that the states which take part in the edge localization are either the rung pair states or two-particle onsite pair state or a superposition of these states. However, for $U=30$,  these bound states no longer remain on the edge rather spreads throughout the lattice as revealed by the correlation matrix plotted in Fig.~\ref{fig:softcore-density_edge}(f). This also reveals that although the density evolution shown in Fig.~\ref{fig:softcore-density_edge}(a) and (c) look similar,  the one in (c) corresponds to the dynamics of the bound states only.

\begin{figure}[t]
	\centering
    \includegraphics[width=1\columnwidth]{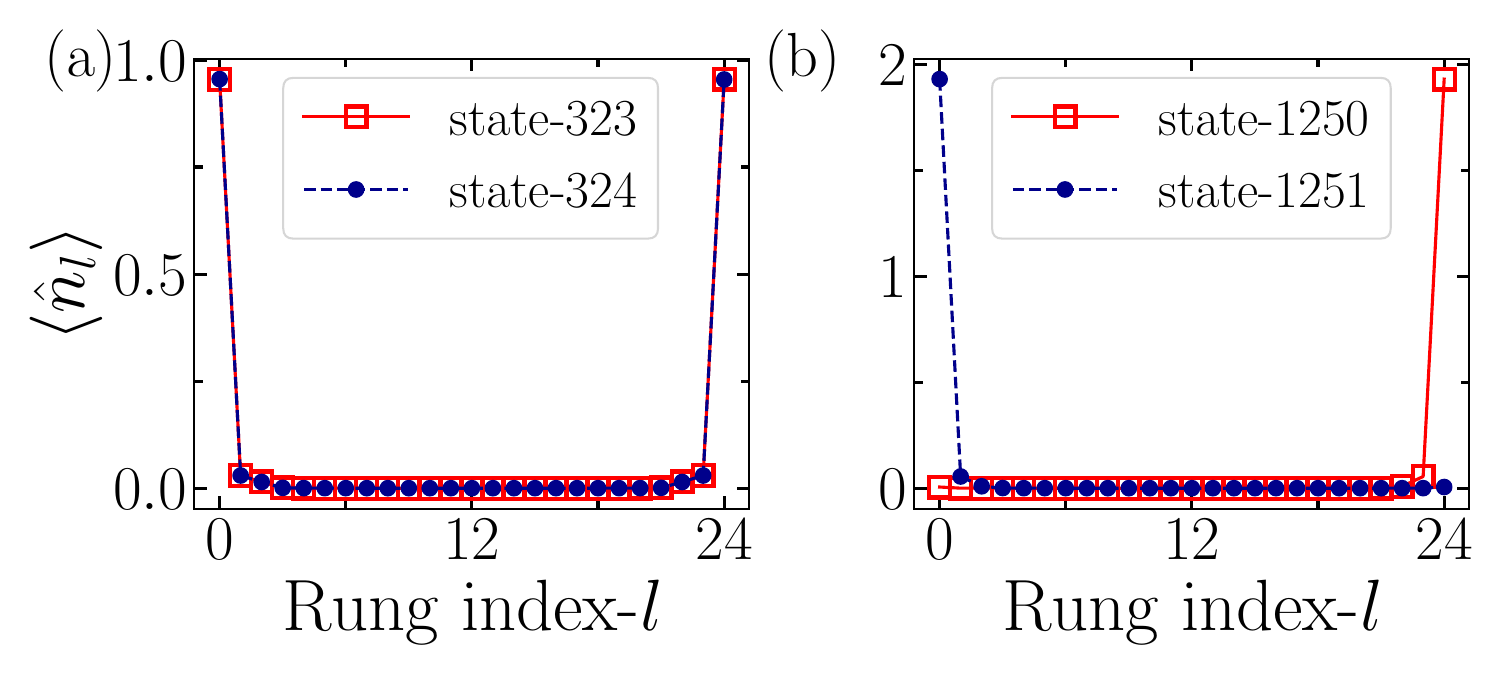}
	\caption{(a) and (b) show the particle density at each rung for the eigenstate with higher IPR values.}
	\label{fig:den_vs_index}
\end{figure}
\begin{figure}[b]
	\centering
	\includegraphics[width=1\columnwidth]{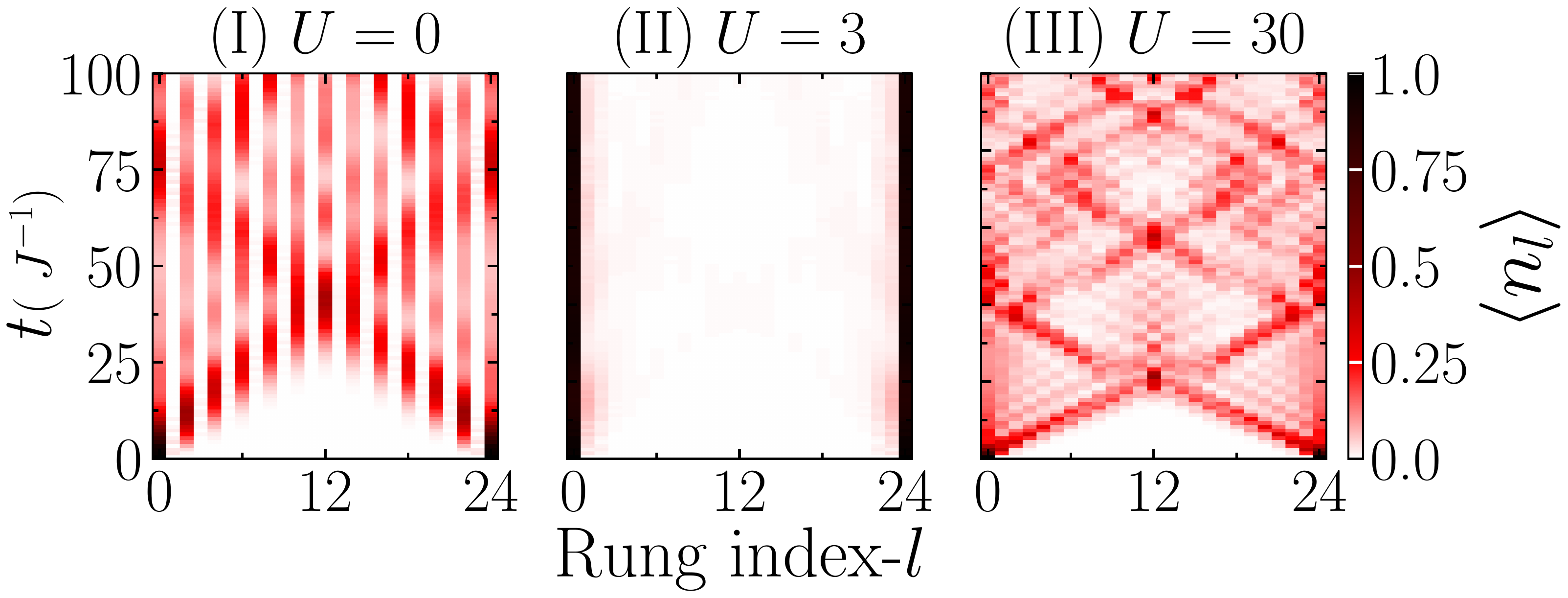}
	\caption{(I-III) Show the density evolution of the two particles with the initial state $|\psi(0)\rangle$ given in Eq.~\ref{eq:ini2} for $U = 0$, $U = 3$, and $U = 30$, respectively and for $\phi=\pi$.}
	\label{fig:softcore-density_two_edge}
\end{figure}
To understand this dynamical edge localization of the bosons we examine the equilibrium energy spectrum. The degree of localization of the states can be quantified from the  
inverse participation ratio (IPR)~\cite{hobbyhorse_2019,Dell_2019} which is defined as 
\begin{equation}
    \rm{IPR}=\sum_i|\psi_i|^4,
\end{equation}
where $\psi_i$ is the $i^{th}$ element of the eigenstate. We plot $\rm{IPR}$ of each state in the spectrum as a function of the eigenenergy and the eigenstate index as shown in Fig.~\ref{fig:Energy_vs_index}. Four localized states appear which are identified as the states possessing the maximum IPR values which are shown in the zoomed in regions in the insets of Fig.~\ref{fig:Energy_vs_index}. The eigenstates with higher IPR values are well isolated from the continuous energy spectrum. These correspond to the states which are localized at the edges of the ladder as can be seen from Fig.~\ref{fig:den_vs_index} where the rung densities are plotted for each state (with higher IPR values). The initial state under consideration in the dynamics which is shown in Eq.~\ref{eq:ini1} corresponds to the state = $1251$(Fig.~\ref{fig:den_vs_index}(b)) and therefore, we observe the bosons edge localize in the dynamics.  From ~\ref{fig:den_vs_index}(a), it is also evident that similar edge localization is possible by considering the initial state as the linear superposition of the states where the particles are localized at the two edge rungs of the ladder. To confirm this behaviour we plot the time evolution of the density $\langle n_l(t)\rangle$ for the initial state 
\begin{equation}
   |\Psi(0)\rangle = \frac{1}{\sqrt{2}}(\hat{b}_{0,A}^{\dagger}\hat{b}_{0,B}^{\dagger}+\hat{b}_{24,A}^{\dagger}\hat{b}_{24,B}^{\dagger})|vac\rangle 
   \label{eq:ini2}
\end{equation}
in Fig.~\ref{fig:softcore-density_two_edge}. As expected, the bosons tend to localize at both the edges of the ladder in the QW when $U\sim 3$. 
\begin{figure}[t]
	\centering
	\includegraphics[width=1\columnwidth]{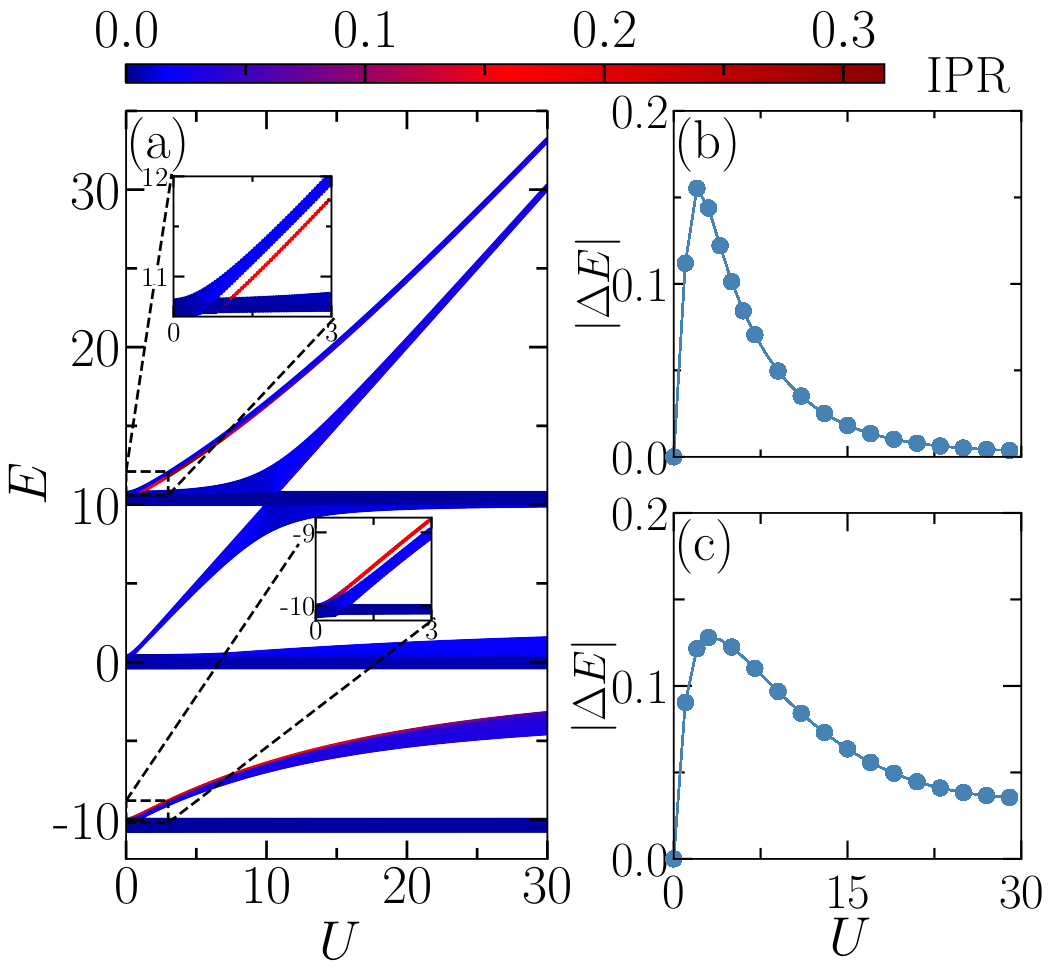}
	\caption{ (a) IPR of the states is plotted as a function of energy eigenvalues $E$ and $U$ for fixed values of rung hopping strength $K=5$ and $\phi=\pi$. The isolated localized states are shown in the insets for clarity. (b) and (c) show the behaviour of $|\Delta E|$ as a function of $U$ corresponding to the localization of particles at one edge and both edges of the ladder respectively. }
	\label{fig:Energy_spectrum_vs_U}
\end{figure}

Furthermore, the edge localization for a range of values of $U$ for fixed $\phi$ can be understood from the energy spectrum as a function of $U$. In Fig.~\ref{fig:Energy_spectrum_vs_U}(a), we plot the IPR of all the states as a function of eigenenergies $E$ and $U$ for $\phi=\pi$. We obtain that for a range of values of $U$, some isolated energy states appear which are identified by their finite IPR (see insets for clarity). These states merge with the continuous energy spectrum when $U$ is not within this particular range. For example, for the localization of states on the left edge of the ladder, this range corresponds to the red region in Fig.~\ref{fig:P_avg_vs_U}. To further quantify this we plot $|\Delta E|= |E_{edge}-E_{bulk}|$ in Fig.~\ref{fig:Energy_spectrum_vs_U}(b,c), where $E_{edge}$ and $E_{bulk}$ are the energies of the states that take part in the localization and the nearest state which is in the continuum band. Due to the appearance of isolated states for a range of values of $U$, $\Delta E$ exhibits a finite bump as a function of $U$ as shown in Fig.~\ref{fig:Energy_spectrum_vs_U}(b) and (c) which correspond to the localization at one of the edges and both the edges respectively. 
\begin{figure}[t!]
	\centering
	\includegraphics[width=1\columnwidth]{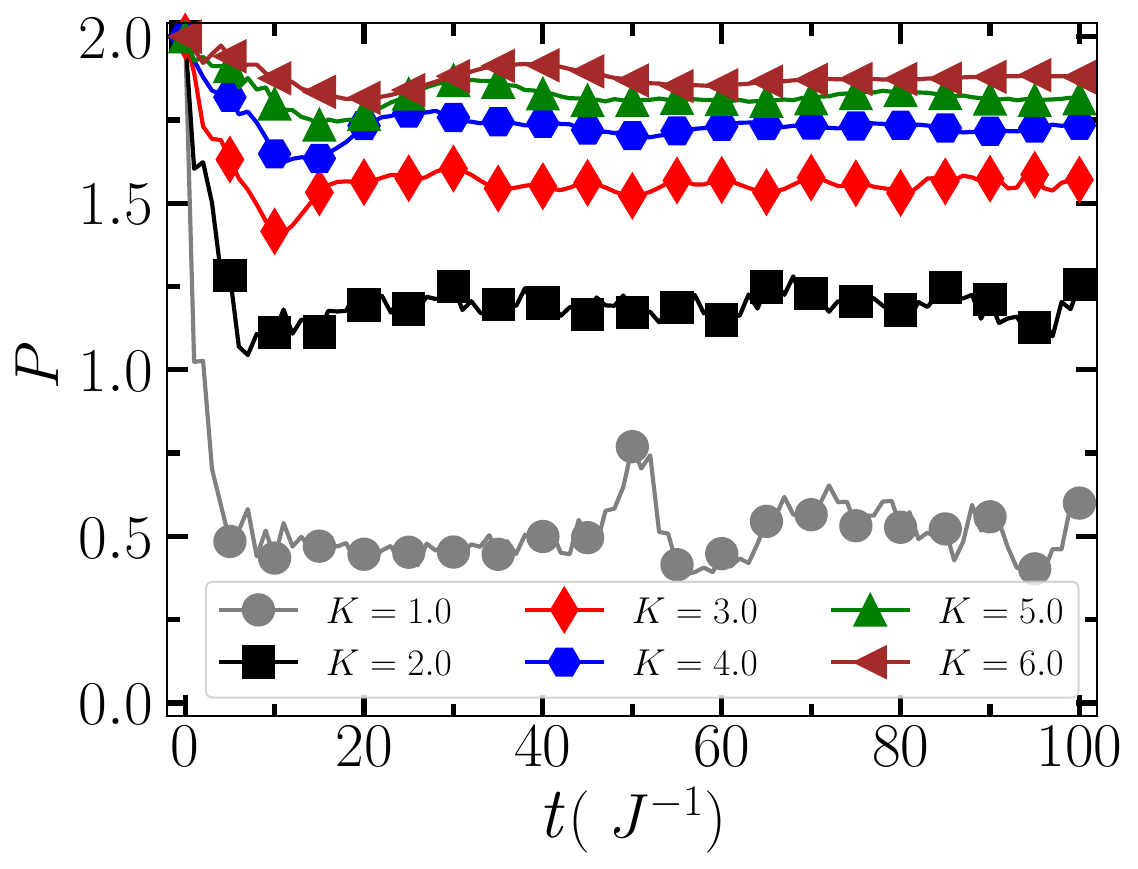}
	\caption{ The figure depicts $P$ as a function of $t(J^{-1})$ for different values of $K$ keeping $U=3$ and $\phi=\pi$.}
	\label{fig:P(t)_Vs_t_vary_K}
\end{figure}

\begin{figure}[b]
	\centering
    \includegraphics[width=1\columnwidth]{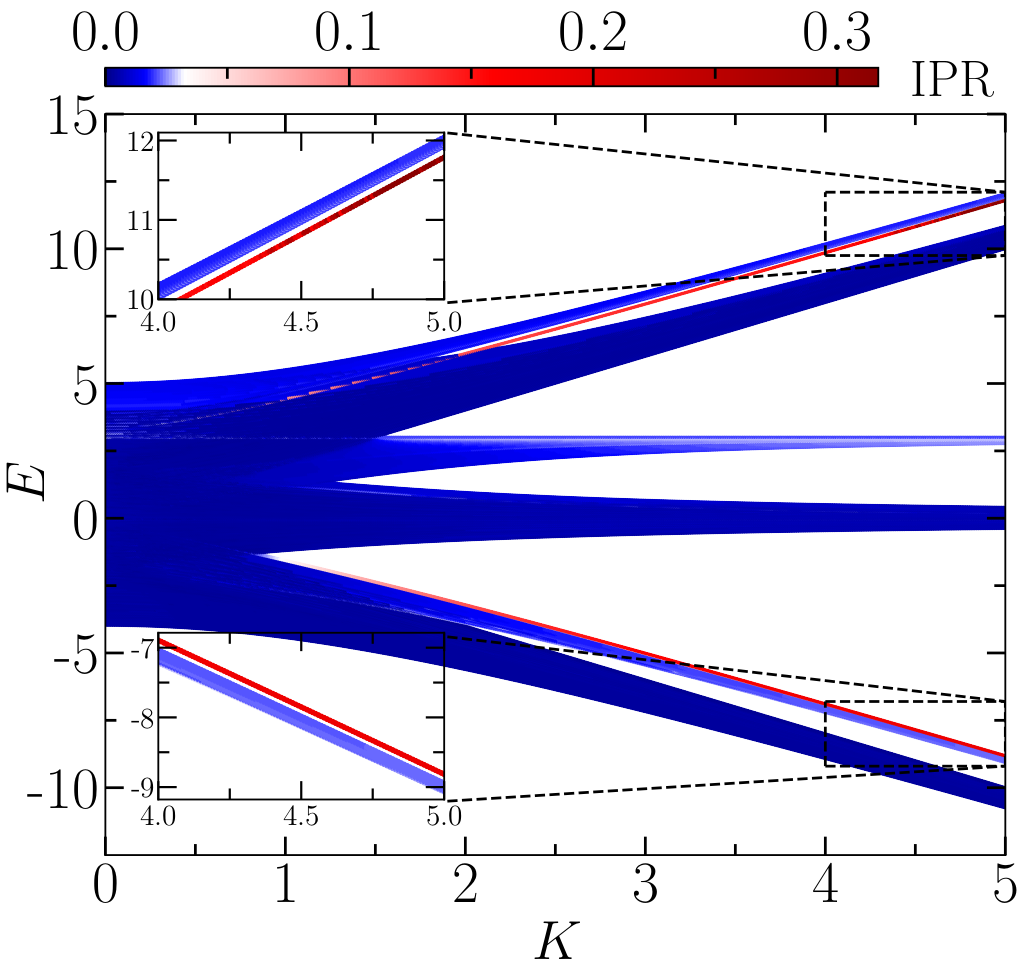}
	\caption{IPR of the states is plotted as a function of energy eigenvalues $E$ and the rung-hopping  $K$  for $U=3$ and $\phi=\pi$.}
	\label{fig:Energy_spec_vs_K}
\end{figure}
We now examine the effect of rung hopping $K$ on the edge localization. To this end we fix $\phi=\pi$ and $U=3$ for which the edge localization is maximum and plot the time evolution of $P$ for different values of $K$ in  Fig.~\ref{fig:P(t)_Vs_t_vary_K}. We obtain that when $K=1$, $P$ saturates to $\sim 0.5$ in the long time evolution indicating that there is no edge localization. However, as $K$ increases, $P$ increases and tends to saturate to a value close to $2$ in the time evolution which is the total number of particles in the system. This suggests that stronger rung-hopping favours edge localization of bosons in the QW. This behaviour can also be confirmed by looking at the IPR of the equilibrium energy spectrum which is plotted in Fig.~\ref{fig:Energy_spec_vs_K} as a function of eigenenergies($E$) and $K$ for $U=3$ and $\phi=\pi$. It can be seen from Fig.~\ref{fig:Energy_spec_vs_K} that for $K < 2$, all the states exhibit smaller IPR $\sim 0$. However,  as $K$ increases, four isolated states emerge with higher IPR values which are already identified as localized states in Fig.~\ref{fig:Energy_vs_index}.

From the above analysis it is revealed that  flux enhances edge localization of softcore bosons in the limit of stronger $K$ and favours a re-entrant delocalization as a function of the onsite interaction strength.  In the following we will study the role of flux on the QW of interacting bosons when they are initialized on the bulk rung of the ladder.

\begin{figure}[t]
	\centering
    \includegraphics[width=1\columnwidth]{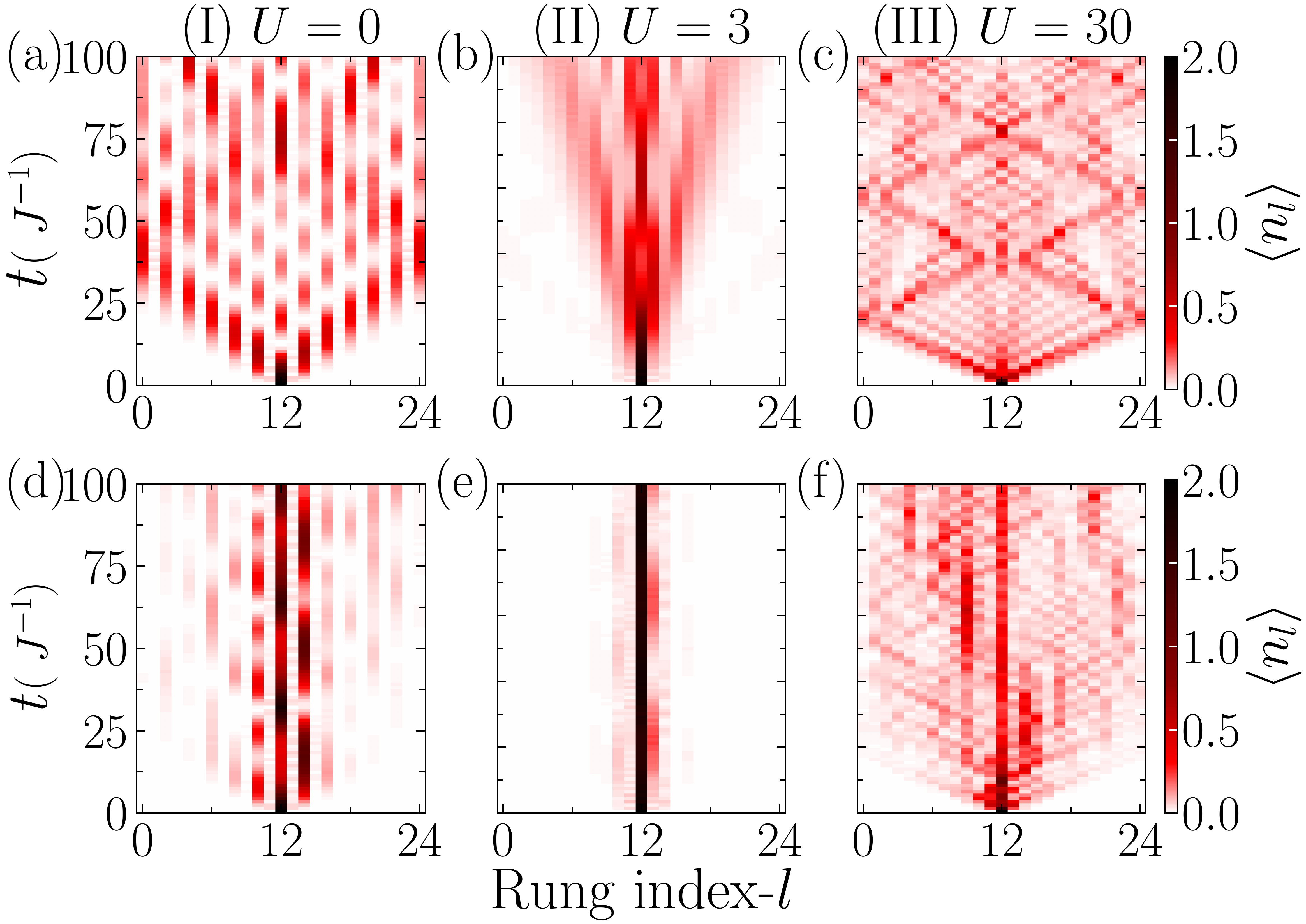}
	\caption{ (I-III) Figure shows the density evolution of the initial state $|\psi(0)\rangle$ given in Eq.~\ref{eq:ini3} for $U=0,~3$ and $30$ respectively and $\phi=\pi$. Fig. (a-c) corresponds to the case when $\lambda=0$ and and Fig. (d-f) corresponds to the case when $\lambda = 0.25$ and $\chi=0$. }
    \label{fig:with_without_dis_den_bulk}
\end{figure}

\subsection{Bulk localization}
In this section we consider an initial state  corresponding to bosons residing on the bulk rung of the ladder which is given as 
\begin{equation}
     |\Psi(0)\rangle= \hat{b}_{12,A}^\dagger \hat{b}_{12,B}^\dagger |vac\rangle.
    \label{eq:ini3}
\end{equation}
The QW of this initial state has already been studied in detail in the presence of both flux and interaction~\cite{giri_paul}. In this case, bosons don't localize in the lattice although a non-monotonous behaviour in the dynamics as a function of $U$ appears which can be seen from Fig.~\ref{fig:with_without_dis_den_bulk}(a-c). From the radial dynamics of the density, it is evident that velocity of propagation initially slows down, reaches a minimum and then increases as a function of $U$ (see ~\cite{giri_paul} for details). Now the question arises, if the slowing down of the radial velocity of interacting bosons due to flux can promote the localization of bosons in the presence of weak disorder? To address this question we investigate the QW of interacting bosons on the flux ladder in the presence of an onsite quasiperiodic disorder. Lattices with quasiperiodic disorder are intermediate to random and clean lattices which exhibit well defined localization transitions in low dimensions. Such localization transitions have been extensively studied in the context of the Aubry-Andr\'e model and its variants for non-interacting ~\cite{aubry1980analyticity,hobbyhorse_2019, Biddle2010, Biddle2011, Li2020, shilpi2021, Padhan2022, Vu2023, Lahini2009, Roati2008, Henrik2018}, as well as interacting particles~\cite{subroto_rev_MBL, Iyer2013, Vedika2017, Xu2019, Paredes2022, huang2023, Schreiber2015, Henrik2017, Jakub2020}.
\begin{figure}[t]
	\centering
    \includegraphics[width=1\columnwidth]{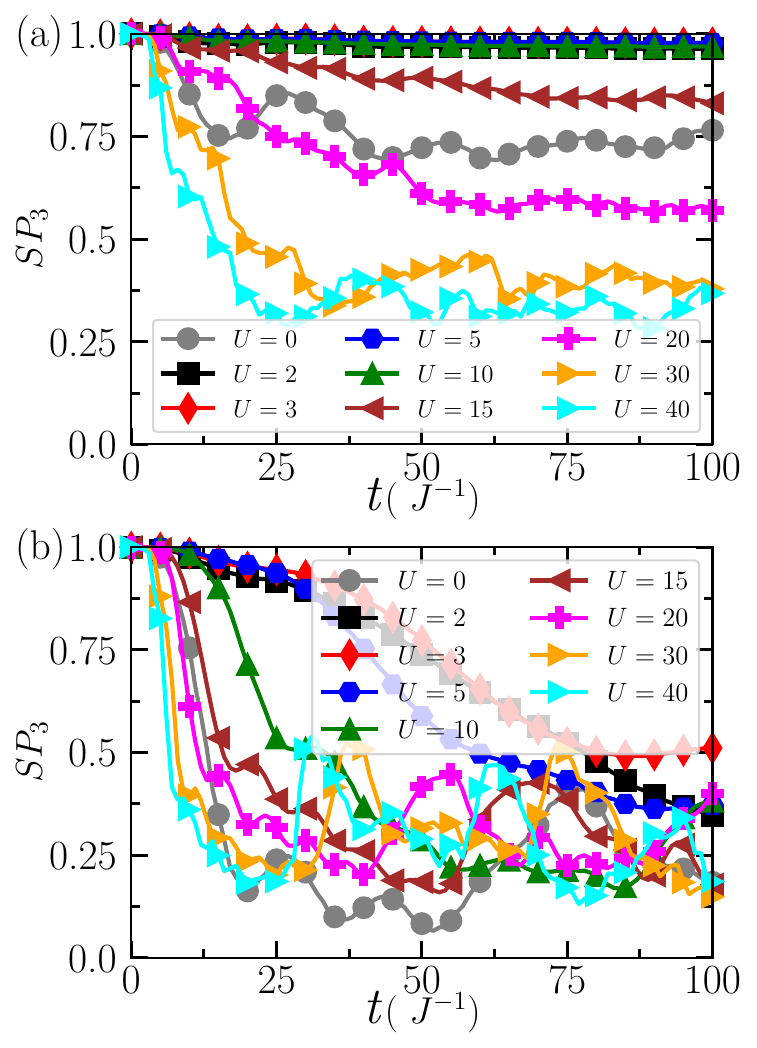}
	\caption{Figures (a) and (b) depict the survival probability $SP_3$ as a function of time $t(J^{-1})$ corresponding to disorder strengths $\lambda=0.25$ and $\lambda=0$, respectively and $\phi=\pi$ with the initial state $|\psi(0)\rangle$ given in Eq.~\ref{eq:ini3}. In (a), the survival probability is computed by averaging over $500$ random $\chi$ values.}
	\label{fig:surviv_prob}
\end{figure}

For our case, we allow quasiperiodic disorder along the legs of the ladder such that the modified Hamiltonian for the system is given as
\begin{equation}
\begin{split}
    H &=  -J\sum_{l,\sigma}(\hat{b}_{l,\sigma}^{\dagger} \hat{b}_{l+1, \sigma}+h.c.)-K\sum_{l} (e^{-i l\phi}\hat{b}_{l,A}^{\dagger} \hat{b}_{l,B}  + h.c)\\ 
    &+\frac{U}{2}\sum_{l,\sigma} \hat{n}_{l,\sigma}(\hat{n}_{l,\sigma}-1)
     +\lambda\sum_{l,\sigma} cos(2\pi\beta l+\chi)\hat{n}_{l,\sigma}, 
\end{split} \label{eqn:ham2}
\end{equation} 
where, $\lambda$ is the strength of the quasiperiodic disorder. We choose $\beta = \frac{\sqrt{5}-1}{2}$ which is the Golden ratio that introduces quasiperiodicity in the lattice and $\chi$ is the phase of the quasiperiodic potential.

\begin{figure}[t!]
	\centering
    \includegraphics[width=1\columnwidth]{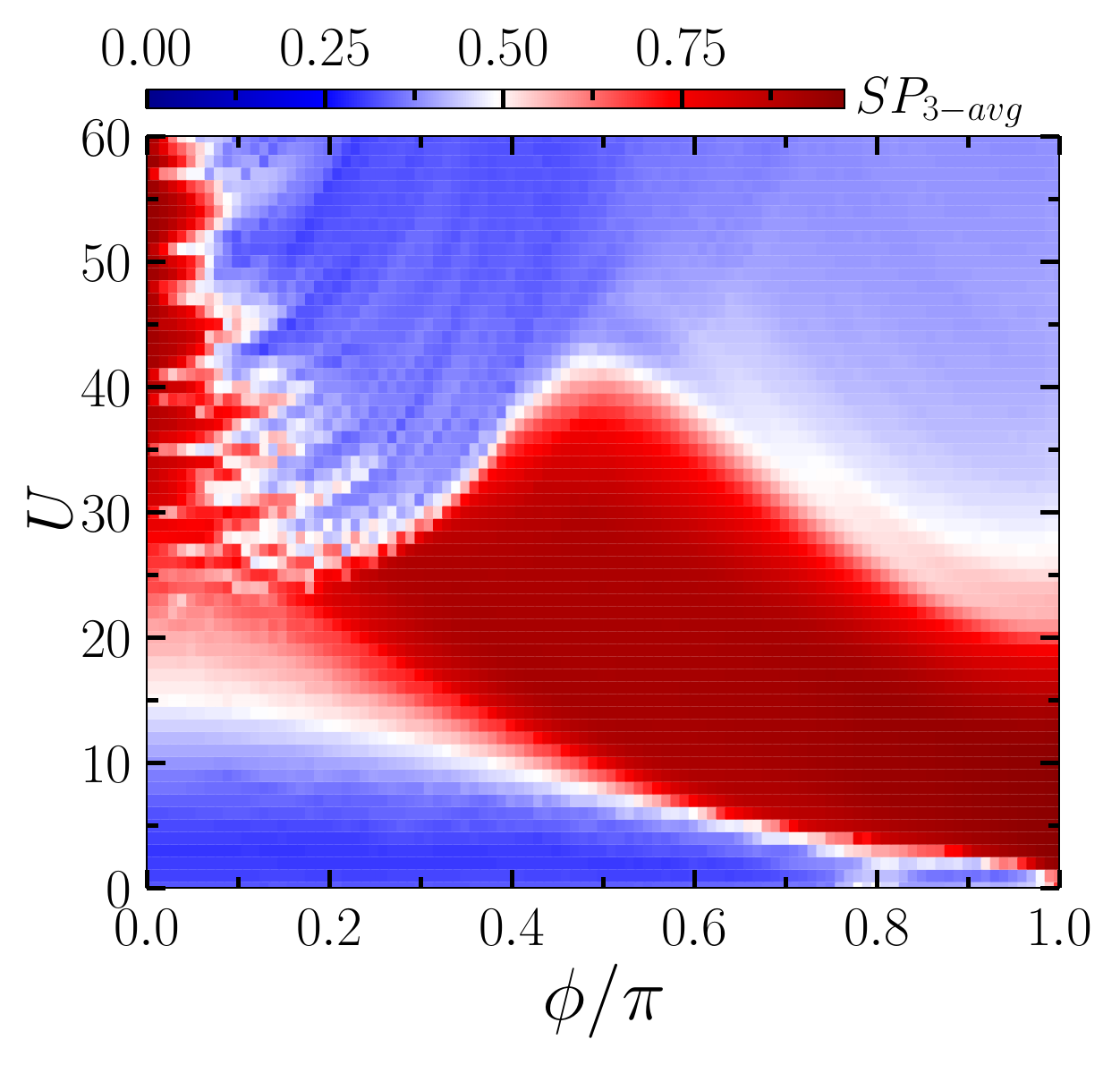}
	\caption{Figure shows $SP_{3-avg}$ as a function of $\phi/\pi$ and $U$ with the  initial state $|\psi(0)\rangle$ given in Eq.~\ref{eq:ini3}. Here the values of $\lambda = 0.25$ and $\chi=0$ are considered and the time average is obtained for time between $t=0(J^{-1})$ to $1000(J^{-1})$ with interval $\Delta t=2(J^{-1})$. }
	\label{fig:with_dis_U_Vs_phi}
\end{figure}

We now investigate the QW by considering a weak disorder strength  in the presence of both flux and interaction. In Fig.~\ref{fig:with_without_dis_den_bulk}(d-f) we plot the density evolution for the values of $U$ considered in Fig.~\ref{fig:with_without_dis_den_bulk}(a-c) for $\phi=\pi$ and $\lambda=0.25$.  
The absence of particle density after a few sites around the initial position in Fig.~\ref{fig:with_without_dis_den_bulk}(e) indicates an absence of linear spreading of density for $U=3$ and is a characteristic feature of dynamics exhibited by localized states. This implies the localization of states for some intermediate value of $U$ (i.e. $U=3$ in this case) where the radial velocity was found to be minimum in the absence of disorder (compare with Fig.~\ref{fig:with_without_dis_den_bulk}(b)). However, such behaviour in the dynamics is not visible for $U=0$ and $U=30$ where the bosons exhibit linear spreading as shown in  Fig.~\ref{fig:with_without_dis_den_bulk}(d) and (f) respectively.

To quantify this localization we compute the survival probability which is defined as 
\begin{equation}
     SP_r(t) = \frac{1}{N}\sum_{l=-r}^{r}\langle \psi(t)| \hat{n}_l |\psi(t)\rangle ,
\end{equation}
where $N$ is the total number of particles. In this case, the quantity $SP_r$ when calculated within a range of sites $r$ around the initial position of the particles tends to approach one if there is localization in the system. In Fig. ~\ref{fig:surviv_prob}(a), we plot $SP_r$ by setting $r=3$ around the central rung as a function of time for different values of $U$ and fixing $\lambda=0.25$ and $\phi=\pi$. 
Starting from $U=0$, the time evolved survival probability first increases reaches a maximum value of $SP_3\sim 1$ for a range of values of $U$ and then gradually decreases as $U$ increases. For comparison, we also plot $SP_3$ as a function of time for different values of $U$ for the case of $\phi=\pi$ and $\lambda=0$ in Fig. ~\ref{fig:surviv_prob}(b). We find that for all the values of $U$,  $SP_3$ always decreases in the presence of weak disorder. Which confirms that the flux enhances the localization of states.

The results from the survival probability infers that the flux enhanced localization also strongly depends on the interaction strength and moreover, the localization is found to happen for a range of intermediate values of $U$. To investigate how this range of $U$ varies with $\phi$, we plot the time averaged survival probability
\begin{equation}
SP_{r-avg} = \frac{1}{m}\sum_{t=0}^T SP_r(t) ,
\end{equation}
where $m$ is the number of time steps. We plot $SP_{r-avg}$ as a function of $\phi$ and $U$ in Fig.~\ref{fig:with_dis_U_Vs_phi} by keeping $\lambda=0.25$ and $r=3$. The deep red region is the range of $U$ where $SP_{3-avg}$ attains it's maximum value is an indication of the localization of the states. The blue regions on either sides of the localized region correspond to the delocalization of the state. 

\begin{figure}[t!]
	\centering
    \includegraphics[width=1\columnwidth]{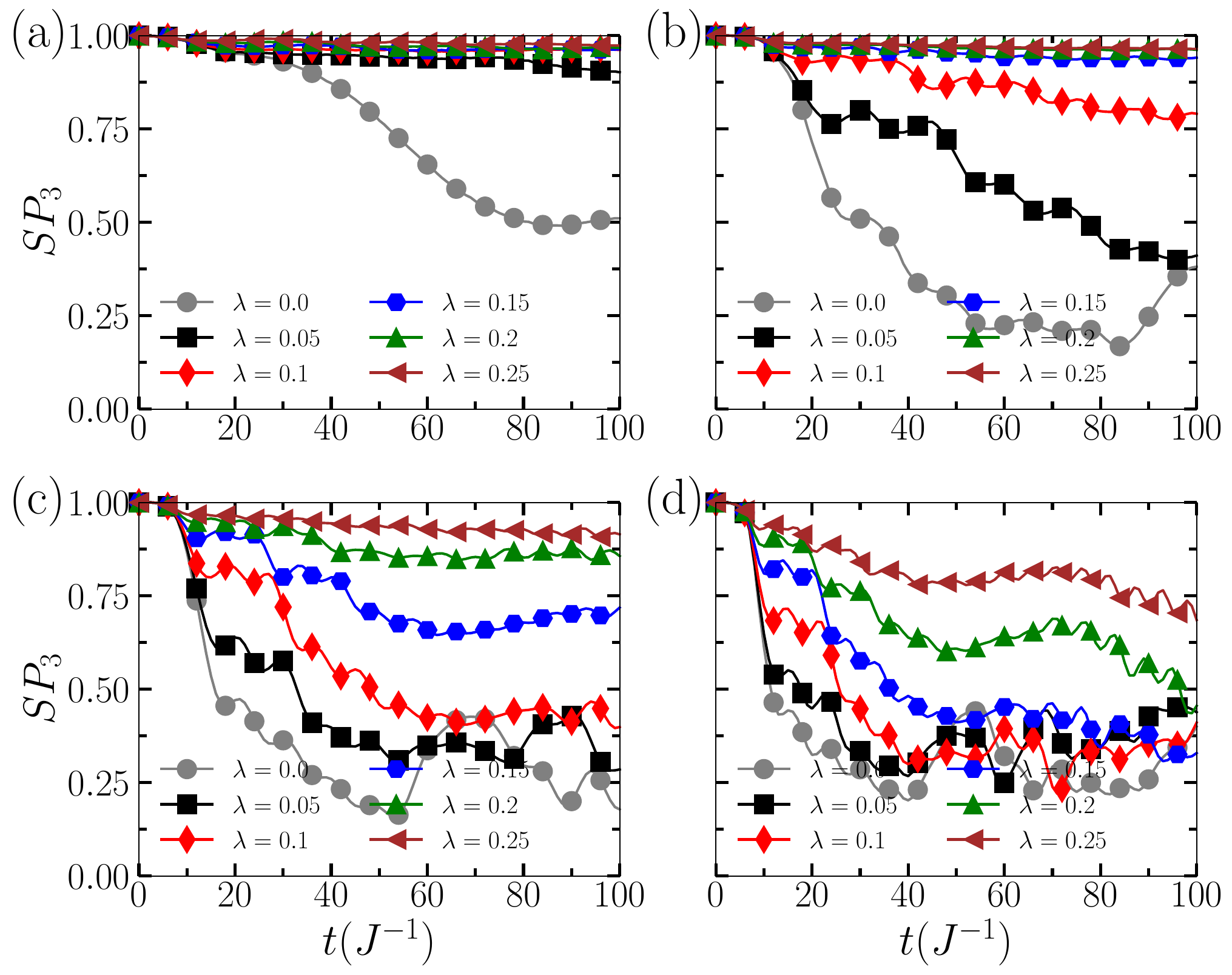}
	\caption{(a-d) show the survival probabilities $SP_3$ as a function of $t(J^{-1}$) for $U=3,~10,~ 15$ and $20$ respectively, with different values of $\lambda$ for each case. Here we start the dynamics from the initial state $|\psi(0)\rangle=\hat{b}_{12,A}^\dagger \hat{b}_{12,B}^\dagger |vac\rangle$ and fix the parameter values $\phi=\pi$ and $\chi=0$. }
	\label{fig:surv_diff_U}
\end{figure}

The value of $\lambda$ for which the  bulk localization occurs is dependent on the values of $U$ and $\phi$. This is shown in Fig.~\ref{fig:surv_diff_U}(a-d), where we plot the survival probability $SP_3$ as a function of $t(1/J)$ for different values of $U$ and $\lambda$ for fixed $\phi=\pi$ and $\chi=0$. It can be seen that when $U=3$, the localization occurs for a very small value of $\lambda\sim 0.05$. However for $U=10$ and $15$, stronger values of $\lambda$ are necessary for the localization to occur. However, for $U=20$, the strength $\lambda=0.25$ is not sufficient for localization to occur. 

\begin{figure}[t!]
	\centering
    \includegraphics[width=1\columnwidth]{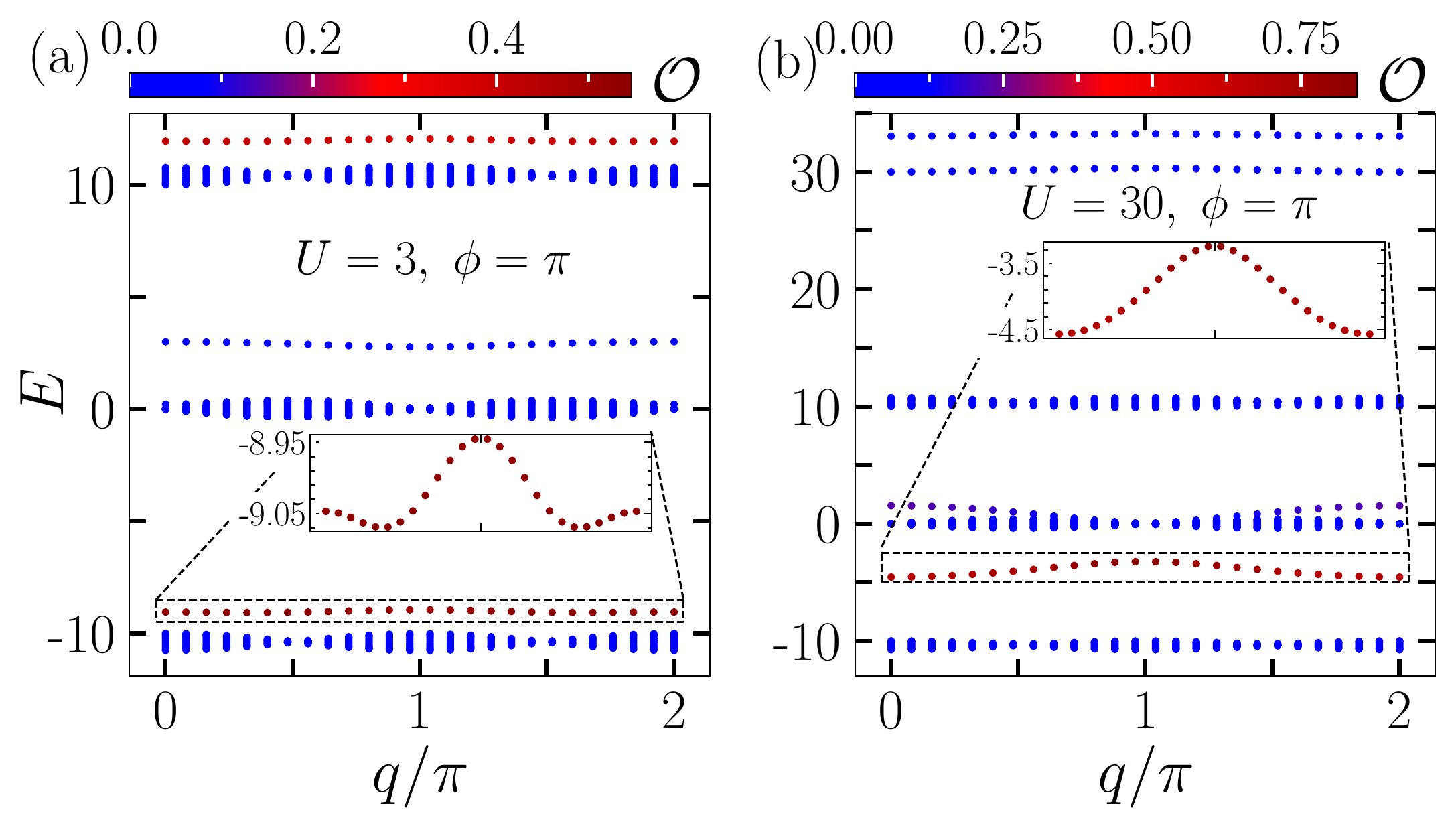}
	\caption{(a) and (b) show the band structures in the absence of disorder for $U=3$ and $U=30$ respectively. In both cases, we consider $\phi=\pi$. The zoomed in data is shown in the insets for clarity.}
	\label{fig:band}
\end{figure}

The re-entrant behaviour of the bulk localization can be understood from the band structure in the absence of $\lambda$ as shown in Fig.~\ref{fig:band}(a) and (b) for $U=3$ and $30$ respectively while fixing the value of $\phi=\pi$. These two values of $U$ correspond to the localization and delocalization scenarios respectively when disorder is present in the system.  We calculate the overlap $\mathcal{O}=|\langle \Phi|\alpha_i\rangle|^2$ between the eigenstates $|\alpha_i\rangle$ and the rung pair state $|\Phi\rangle$ (which is our initial state) of the momentum space Hamiltonian and plot them as color coded data in Fig.~\ref{fig:band}(a) and (b) for all the states. In this case, the eigenstates that exhibit non-zero values of $\mathcal{O}$ contribute to the dynamics. It can be seen that the width of the band that contributes to the dynamics is much flatter for $U=3$ than for $U=30$ (compare Fig.~\ref{fig:band}(a) with (b)). The smaller bandwidth indicates the lower effective hopping($J_{eff}$) and hence slowing down of the particles. As the localization of the system depends on the ratio $\lambda/J_{eff}$, a smaller value of $\lambda$ favours localization for some particular values of $U$ and $\phi$.

These findings reveals that for sufficiently weak disorder strength, flux favours localization of the quantum states of two interacting bosons. The localization also strongly depends on the interaction strength where for a fixed value of $\phi$ the system undergoes delocalization to localization and then to delocalization transition as a function of interaction strength. This phenomenon of delocalization to localization and to delocalization in the interacting system can be termed as a signature of the re-entrant delocalization transition in the interacting system.

\section{Conclusion}
In this work, we have investigated the quench dynamics of two interacting bosons in a two-leg ladder under the influence of an artificial gauge field. Our investigation reveals two interesting scenarios in the context of localization of interacting particles in the simplest possible system. In the first case, we have shown that when the two bosons are initialized on the edge rung of the ladder, the interplay of flux and finite onsite interaction results in an edge localization of the bosons. We have obtained that for a fixed value of flux, such edge localization occurs for a range of values of interaction strengths. Moreover, this range is found to be dependent on the flux strength i.e. the localization occurs for stronger interaction strength when the flux strength is weaker and vice-versa. On the other hand we have found that in the case of bosons initialized on the bulk rung, such localization does not happen. However, in the later case, if a sufficiently weak quasi-periodic disorder strength is introduced, a clear localization is found to be favoured by the flux. In both the cases, the localization is favored when the rung hopping is stronger than the leg hopping in the ladder. This flux enhanced localization leads to a feature of re-entrant delocalization where the delocalized states first localize and then delocalize again. 

The quench dynamics on flux ladders have been experimentally studied recently in system of ultracold atoms in  optical lattices~\cite{Gadway2017direct,science.aaa8515,Gadway2018a,Gadway2018b,Atala2014,Tai_2017, Li2023, tao2023}. For example, Ref.~\cite{Tai_2017} explores the quench dynamics of two interacting bosons on a system described by the model considered in our analysis. Hence, our finding can be immediately be observed in the experiments using artificial systems. On the other hand our study provides a mechanism for localization of interacting bosons in the absence of disorder or in the presence of weak disorder at the two particle level which poses open questions on the stability of such phenomena in the true many-body limit, in the presence of strong disorder and off-site interactions. 

\section{Acknowledgement}
T.M. acknowledges support from Science and Engineering Research Board (SERB), Govt. of India, through project No. MTR/2022/000382 and STR/2022/000023.

\bibliography{reference_new}
\end{document}